# Sperm-hybrid micromotor for drug delivery in the female reproductive tract

*Haifeng Xu[1], Mariana Medina-Sánchez[1]\*, Veronika Magdanz[1], Lukas Schwarz[1], Franziska Hebenstreit[1] and Oliver G. Schmidt[1,2]*

[1]Institute for Integrative Nanosciences, IFW Dresden, Helmholtzstraße 20, 01069 Dresden, Germany

[2]Material Systems for Nanoelectronics, Chemnitz University of Technology, Reichenhainer Straße 70, 09107 Chemnitz, Germany

ABSTRACT

A sperm-driven micromotor is presented as cargo-delivery system for the treatment of gynecological cancers. This particular hybrid micromotor is appealing to treat diseases in the female reproductive tract, the physiological environment that sperm cells are naturally adapted to swim in. Here, the single sperm cell serves as an active drug carrier and as driving force, taking advantage of its swimming capability, while a laser-printed microstructure coated with a nanometric layer of iron is used to guide and release the sperm in the desired area by an external magnet and structurally imposed mechanical actuation, respectively. The printed tubular microstructure features four arms which release the drug-loaded sperm cell *in situ* when they bend upon pushing against a tumor spheroid, resulting in the drug delivery, which occurs when the sperm squeezes through the cancer cells and fuses with cell membrane. Sperms also offer higher drug encapsulation capability and carrying stability compared to other nano and microcarriers, minimizing toxic effects and unwanted drug accumulation. Moreover, sperms neither express pathogenic proteins nor proliferate to form undesirable colonies, unlike other cells or microorganisms do, making this bio-hybrid system a unique and biocompatible cargo delivery platform for various biomedical applications, especially in gynecological healthcare.

INTRODUCTION

The development of drug delivery systems that provide effective doses locally in a controlled way is one of the main challenges in the worldwide fight against cancer.[1, 2] Among such nano- and microcariers, liposomes were shown to be able to lower the cytotoxicity, improve the solubility and control the release of drugs.[3] Despite surface functionalization that allows them to increase the circulation time[4] and active tumor targeting ability,[5] there are still a lot of challenges to address, such as unspecific uptake by other organs,[6] limited tissue penetration[7] and the decrease of effective concentration due to the dilution in body fluids.[8] To overcome the abovementioned challenges, novel cellular carriers have been proposed, that take advantage of the membrane fluidity, long lifespan and high biocompatibility of cells.[9] As examples, stem cells as a combinatorial drug delivery system have been used for regenerative therapy,[10] macrophages for anticancer drug delivery[11] and red blood cells for sustained drug release in blood.[12] However, a precise transport method is particularly necessary for targeted drug delivery. Self-propelled cells, which provide a combination of cellular encapsulation and propulsion, are attractive due to their biocompatibility, ability to interact with other cells/tissue and optimal swimming performance in physiological microenvironments.[13, 14] Bacteria, for example, have been reported as promising self-propelled carriers.[15] Combined with their chemotactic,[16] and/or magnetotactic properties[17] or associated guidance components,[18] bacteria can transport and deliver drugs or perform *in situ* sensing. It is however noteworthy that rapid clearance or even autoimmune reactions might be caused by the immune response to certain bacteria.[19]

In this work, we report a new type of sperm-hybrid micromotor and its potential application in targeted drug delivery to treat gynecologic cancers, including cervical, ovarian, uterine, vaginal, vulvar and fallopian tube cancers, which are affecting more than 100.000 women every year in the United States.[20] The here reported micromotor comprises a motile sperm that serves as the propulsion source and drug carrier, and a 3D printed four-armed microtube, also called "tetrapod", used for the magnetic guidance and mechanical release (Figure 1a). Compared to other drug carriers, sperms are naturally optimized to swim efficiently through the female reproductive system and can load a high amount of drugs,[21] making them good candidates as drug carriers toward gynecologic cancer treatment. The sperm cell provides an extraordinary ability to encapsulate protein and other hydrophilic drugs owing to its crystalline nucleus.[22] The sperm membrane can protect drugs from body fluid dilution, immune-reactions and the degradation by enzymes. Sperms, with their compact membrane system, can efficiently avoid dose dumping, which is regarded as a major issue of micelles.[23] For microscale carriers such as microspheres, micro capsules, and drug-loaded micromotors,[24, 25] drug uptake is always a problematic issue.[26]

Sperm cells, with their somatic cell-fusion ability,[27] are expected to improve the drug transfer to the target cells and the drug availability as well. Several proteins from the sperm membrane, such as CD9 and integrins are involved in this process.[27] It solves the uptake problem of microscale carriers via the nanoscale mechanism of membrane fusion. Providing controllable guiding and release mechanisms, sperm-hybrid micromotors can potentially deliver drugs precisely to tumor tissue and furthermore avoid undesired drug accumulation in healthy tissue. The concept of a hybrid microdevice consisting of a sperm cell and a synthetic component was introduced in our previous research, where micromotors have been successfully employed to transport and guide single sperm cells as a promising approach for *in vivo* assisted fertilization.[13, 28] In the here presented work, we pursue a different application, however staying in the field of gynecologic medicine to take advantage of the sperms' natural adaptability to the reproductive system. This novel system combines several intriguing features, namely self-propulsion, *in situ* mechanical release of the drug-loaded sperm, penetration ability and high drug loading capacity. Besides, sperm-hybrid micromotors are exceptionally well-suited to swim through the female reproductive tract due to the naturally developed specialization of the cellular component.

RESULTS

*Guidance and sperm release*

The tetrapod microstructure was designed to have a tubular body and four flexible arched arms (Figure 1a). These arms protrude from one opening of the microtube in a curved manner. The dimensions of the microstructure are shown in Figure 1b,c. At the narrowest point between the four arms, the maximum distance is 4.3 µm. In preliminary experiments, the dimensions of the structure were optimized according to the dimensions of a sperm cell (for the experiments shown in this manuscript, bovine sperm cells were selected for their similar paddle-like shape to human sperms), of which the head is on average 4.5 µm wide, 1 µm thick and 10 µm long.[29, 30] Therefore, the sperm can be captured to propel the microstructure. Once the arms hit a substantial barrier, such as a cell cluster, they bend to enlarge the distance between the arms allowing the sperm to escape from the tube in the process (Figure 1a). The polymeric structure was designed and fabricated by means of two-photon 3D nanolithography (Figure 1c). Then the tetrapod microstructure was asymmetrically coated with 10 nm of iron with a tilt angle of 15º to create a magnetic "easy axis". An additional layer of 2 nm of titanium was deposited to improve the composite's biocompatibility. A simulation was performed with finite element software to validate the optimum tetrapod geometry (detailed parameters shown in SI). This simulation indicated that the applied force of 128 pN by a motile sperm[31] results in the arms deformation, increasing the aperture or diagonal distance between them in about 328 nm (Figure 1d). Such deformation is enough for the sperm to be released. The enlargement would be increased up to 1018 nm when the applied force is 450 pN which could be generated by a hyperactivated sperm.[32] To ensure the sperm release from the microstructure, progesterone was added to the sperm medium to hyperactivate sperms.[33] Video S1 demonstrates the elasticity and shape memory of the arms by repeatedly pressing down the arm with an AFM tip towards the substrate. The arms were bent but returned to their original position without damage.

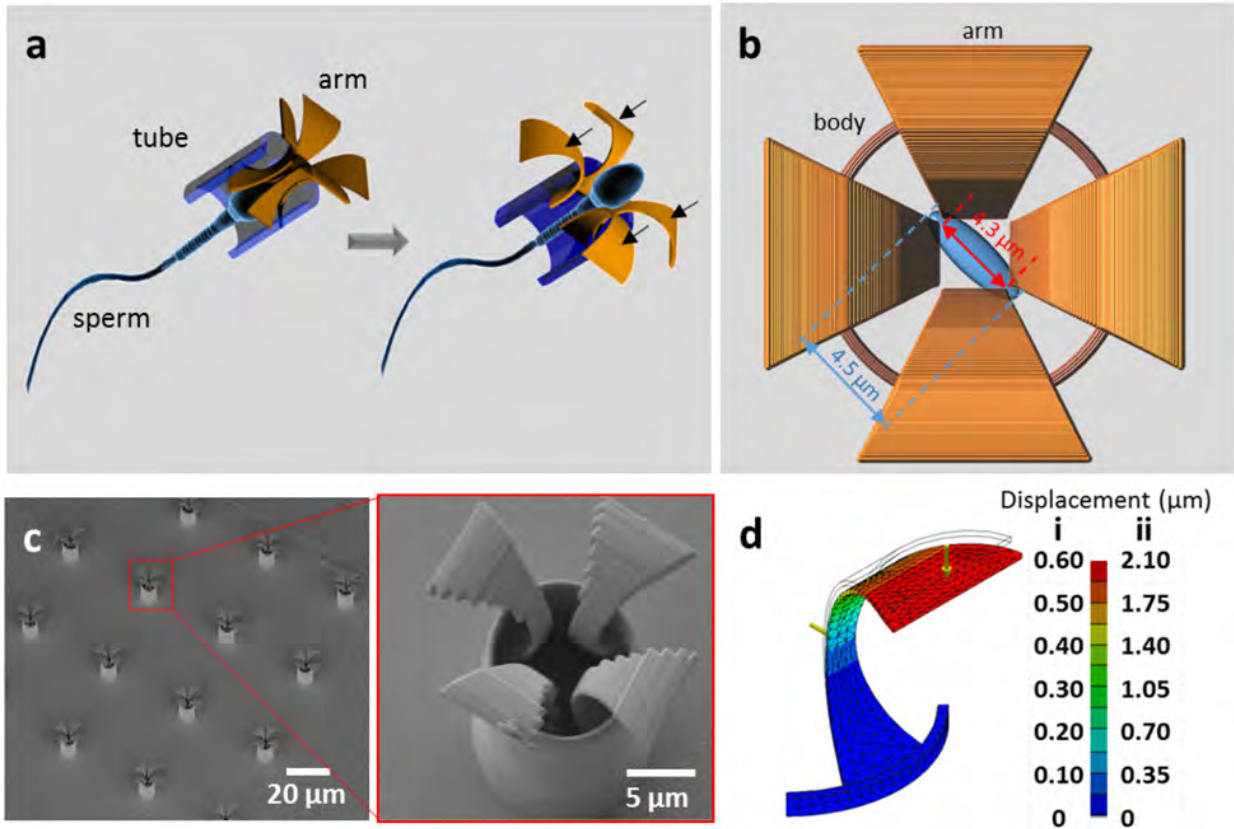

**Figure 1.** (a) Schematic illustration of the sperm-hybrid micromotor and the sperm release process. Black arrows represent the reactive force on the arms upon hitting an obstacle. (b) Top view of the tetrapod microstructure with schematic sperm head (c) SEM images of an array of printed tetrapod microstructures. (d) Simulation results demonstrating the deformation of one single arm. Yellow arrows represent the applied forces. (i) Applied force is 128 pN from a motile, non-hyperactivated sperm. (ii) Applied force is 450pN from a hyperactivated sperm.

When an approaching sperm reaches the microstructure, it gets mechanically trapped inside the cavity of the tubular part and start to push the tetrapod forward (Video S2). The tubular body of the tetrapod is only 2 µm longer than the sperm head, thus the sperm tail can still beat freely outside the tube to provide powerful propulsion as it was previously demonstrated by our group.[34] Compared to free sperms, the average swimming velocity of the sperm-hybrid micromotors is nonetheless decreased by 43% from $73 \pm 16$ µm/s to $41 \pm 10$ µm/s (for 15 samples of sperm-hybrid micromotors). The main reason for the velocity reduction is thought to be the increase of the fluid drag that is provoked by the synthetic material and the complex structure of the tetrapod. It was reported that the swimming velocity of sperms is influenced by temperature and the rheology of the medium as well.[35] The asymmetrically distributed metal coating makes it possible to guide a tetrapod microstructure or a sperm-hybrid micromotor and even manipulate several of them simultaneously (Video S3). Figure 2a illustrates a rectangular track of a guided sperm hybrid-micromotor, also shown in Video S3. The hybrid motor was easily steered by changing the direction of the external magnet. In most cases, the hybrid micromotor rotates while it moves forward due to the helical motion of the sperm,[36] which means the tetrapod does not change the characteristic motion of the sperm. Figure 2b illustrates vertical guidance of a sperm-tetrapod. The hybrid micromotor was steered to swim vertically out of plane simply when the external magnet was placed vertically. Thereby the swimming depth of the tetrapod could be adjusted. Pluronic® F-127 solution was reported previously as sperm repellent agent.[37] Prior to the experiments with sperm cells, tetrapods were thus treated with Pluronic® F-127 solution to reduce undesired adhesion between the sperm membrane and the tetrapod surface. Video S4a shows the motion of the sperm-tetrapod without adhesion in which the sperm cell rotates inside the tetrapod. In the non-adhesion situation, sperms immediately swam out when the tetrapod arms hit an obstacle. However, the majority of tetrapods was found to rotate together with captured sperms due to either surface interaction between the sperm membrane and the material surface or the mechanical locking of the sperm head inside the structure (Figure 2c). In these cases, it always took several seconds from the moment the tetrapod hit a wall until complete release of the contained sperm (Video S4b,c).

PDMS microfluidic channels were fabricated as a platform for the investigation of the sperm release mechanics. Sperm release occurred when two arms hit a corner (Video S4a,b) or when four arms hit a wall (Video S4c), while the release processes are different in both cases. When the motors came into contact with the targeted barriers, they still rotated for a while after the forward swimming was stopped. The rotating sperm-tetrapod stopped faster when two arms hit a corner because of the geometric gap between the arms which was easily caught on a corner. Once the rotation stopped, the sperm cell escaped when the tetrapod arms opened (release in 7 s). When four arms hit a wall, the rotation was not stopped because the arms were not locked. Thus, the sperm release took longer when four arms were bent on a wall (12 s). In both cases, tetrapods were pushed back by around 3 µm after the sperms escaped. The reason for these recoils is the existence of an elastic force that makes the tetrapod arms recover their original shape once the pushing sperm is gone. Even though there is a substantial diversity in bovine sperm dimensions, swimming behaviors and fabricated tetrapods within a sample, more than 2/3 (15 out of 22) of the coupled motors were shown to successfully release sperm cells.

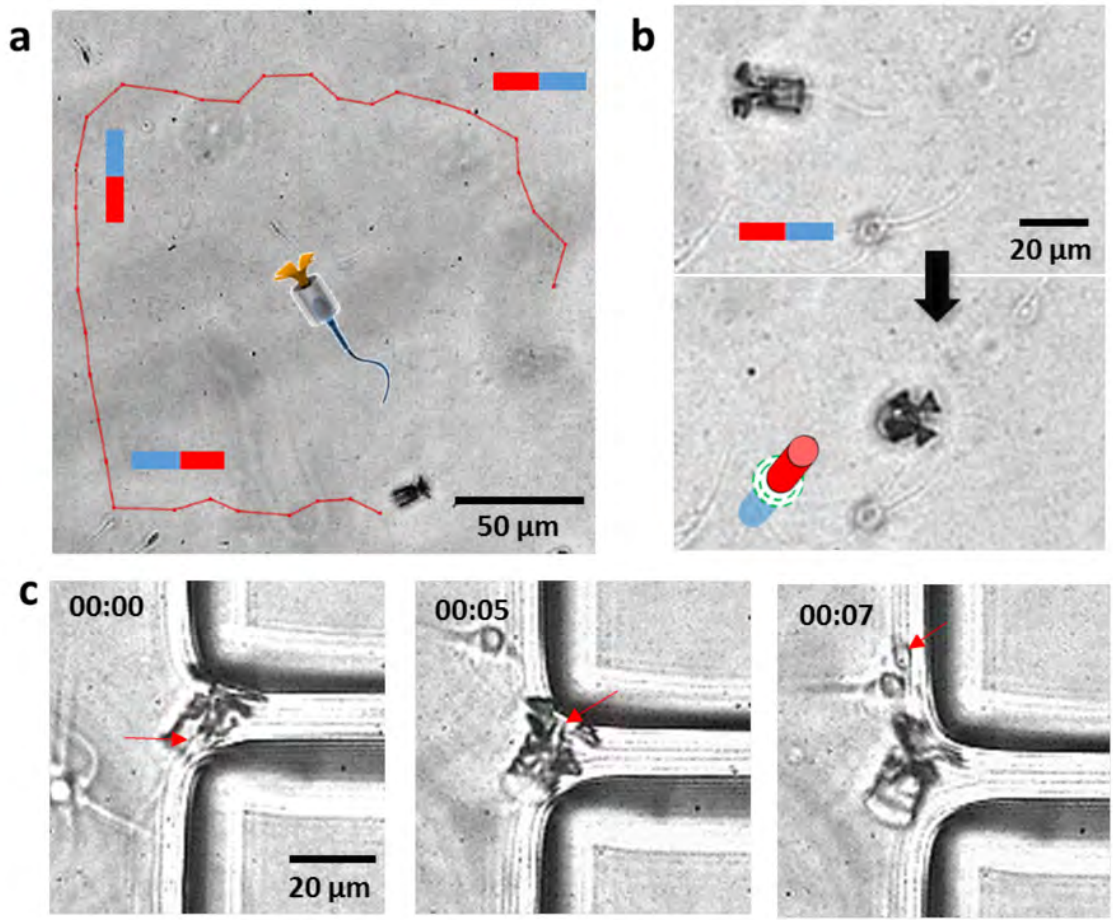

**Figure 2.** (a) Rectangular track (red line) of a sperm-hybrid micromotor under magnetic guidance in the horizontal plane; (b) Steering of a sperm-hybrid micromotor into the vertical plane. (c) Image sequence of a sperm release process when the arms hit the corner of a PDMS wall. Red arrows point at the sperm head. Time lapse in min:sec.

Previous research has reported similar elastic structures on the microscale that were also fabricated by two-photon lithography.[38] For example, a microscale beam was deformed by around 5 µm under compression with 68 pN.[39] The elasticity of such photosensitive polymeric material depends on the type of monomer and the cross-linking parameters.[38] In order to avoid uncontrolled release events, we did not choose a softer polymer material but optimized the laser power that initializes cross-linking (5 mW). In our simulation, the applied force was given according to the maximum pushing force of a sperm in low-viscosity fluid (2.29 × 10$^{-3}$ Pa·s).[31] It has been reported that the sperm can generate a more powerful force when the head is pushing against an obstacle.[40] Furthermore, the force can be up to 20 times higher when the sperm is hyperactivated and swims in the viscoelastic fluid of the female reproductive system.[32] Consequently, we believe that sperms can be released with such a mechanical trigger system also, and maybe even more efficiently, under physiological conditions.

*Drug loading of sperm*

We used DOX-HCl as a model drug to evaluate the encapsulating performance of sperms. Doxorubicin (DOX) has been approved as a chemotherapy medication with a broad application spectrum in cancer therapy since 1974.[41] Its liposomal form (Doxil) is also used primarily for gynecological cancer treatment.[42] DOX-HCl-loaded sperms are obtained by simple co-incubation of DOX-HCl and live sperms. After purification by centrifugation, the incubated sperm sample can be redispersed in sperm medium (Figure 3a). The fluorescence image in Figure 3b(i) shows that the majority of the sperm cells have been loaded with DOX-HCl (self-fluorescent at 470 nm excitation wavelength), demonstrating an efficiency of 98% with a count of 3502 sperm cells. A 3D reconstruction from a z-stack of images of a single sperm cell is shown in Figure 3b(ii). DOX-HCl was predominately found in the head and the midpiece of the sperm. The cross-section of the sperm in Figure 3b(iii) indicates that a larger amount of DOX-HCl was loaded in the cytoplasm and the nucleus as compared to the cell membrane. Drug loading efficiency was evaluated by calculating the loading ratio. The drug loading amount was determined by the difference between the initial amount of DOX-HCl before incubation and the residual amount in the supernatant after co-incubation, which were both quantify by their respective fluorescence signals. Figure 3c depicts the drug loading profiles related to the DOX-HCl concentration. In the solution with a concentration of $3\times10^6$ sperms per mL, the loading amount of DOX-HCl increased approximatively linearly with the concentration of DOX-HCl ranging from 10 to 200 µg/mL. Hence, the loading ratio remains at around 15 % for all concentrations. For the maximum concentration of DOX-HCl in our experiments, the loading amount was up to 37 µg/mL in 500 µL sperm solution. This amount indicates an average encapsulation of 15 pg of DOX-HCl per single sperm cell. We also carried out a test on the encapsulating stability. The result shows that only 18 % of drug was leaked into the solvent after 96 h (Figure S2). This means that the DOX-HCl encapsulation by sperm is sufficiently stable for subsequent drug delivery experiments.

The sperm membrane consists of a lipid bilayer with embedded proteins. Sperms are expected to encapsulate most of the hydrophobic drugs into the cytoplasm and the nucleus.[43] After maturation, reduced cytoplasm remains in the midpiece and the head of a healthy sperm.[44] In addition, the condensed chromosomes can absorb a specific amount of hydrophilic substance by binding to proteasomes.[45] After a certain time of co-incubation with sperms, most of the DOX-HCl and BSA (Bovine Serum Albumin, used as a model for protein drugs) was found in the sperm head and the midpiece (Figure 3b and Figure S3). In contrast, BSA was also retained in the sperm tail, which is the result of the protein absorbability of the cell membrane. DOX-HCl attacks cancer cells by interfering with the macromolecular biosynthesis.[46] Unlike cancer cells, mature sperms have terminated most of their macromolecular synthesis due to the lack of the complete endomembrane system, which avoids the digestion of the drug inside the sperm cell.[47] What is more, it is not necessary for sperm cells to express new proteins to maintain their motility. The propelling power is generated in the mitochondria in the midpiece.[44] Since DOX-HCl also interferes with the macromolecular synthesis in mitochondria,[48] it may also inhibit the energy metabolism of sperms, which is probably the reason for the observed decrease of the sperm's average swimming velocity from 71 µm/s to 57 µm/s, which is, however, still an acceptable velocity to serve the sperms' purpose. The result of the drug loading ratio calculation highlights the high encapsulation capacity of sperm cells. Previous research showed successful loading of DOX into macrophages.[49] It was demonstrated that a more concentrated dose of DOX was found inside the cell (2.5 pg per cell), particularly in the nucleus, compared to the DOX solution that was co-incubated with the macrophages (cell size roughly 15-20 µm). We observed a similar phenomenon during the drug loading into sperm cells, as depicted in Figure 3. ATP-mediated permeabilization of the cell membrane was regarded as an explanation for this loading profile. Sperm cells exhibit improved uptake of ionic DOX-HCl[50] compared to molecular DOX, which is normally taken in by facilitated diffusion.[41] Further research is required to investigate the transport mechanism through the sperm's cell membrane. We believe that this approach of using living cells could lead to a new dosage strategy that allows high local doses of anticancer agents while reducing systemic toxic effects. Another advantage of the sperms comes from their incomplete metabolic system.[44] A Sperm cell can protect a contained drug within its lipid bilayer like a liposome, but does not metabolize the drug like stem cells or other somatic cells would do. Because of our observation that the swimming velocity significantly decreased when the DOX-HCl concentration was higher than 100 µg/ml, a concentration of 100 µg/mL was used to prepare DOX-HCl-loaded sperms in the subsequent experiments.

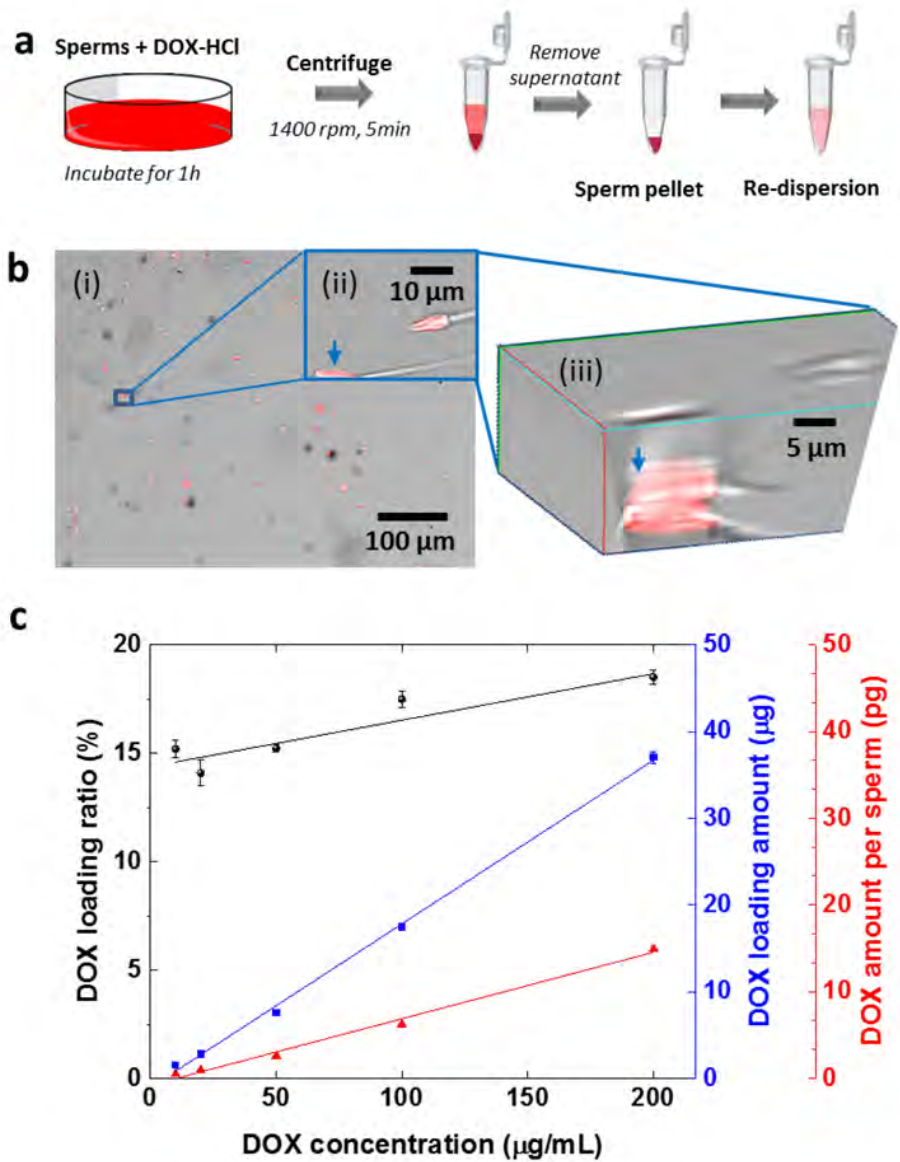

**Figure 3.** (a) Experimental flow chart for loading DOX-HCl into sperms. (b) Fluorescence and brightfield overlay images of DOX-HCl-loaded sperms in (i) 10×; (ii) 40×; (iii) 3D reconstruction of 36 z-stack images with stack separation distance of 0.3 µm (c) Plots of drug loading results versus DOX-HCl concentrations. Drug loading ratio is the ratio of encapsulated DOX-HCl to the original amount of DOX-HCl. Drug loading amount is the encapsulated amount of DOX-HCl in 500 µl sperm solution at a concentration of $3·10^6$ sperms/mL.

*Sperm release and drug distribution in tumor spheroids*

Drug-loaded sperms were first tested on adherent HeLa cells and then on HeLa spheroids with two different model drugs (FITC-BSA and DOX-HCL). Figure S5 illustrates the morphology change of the adherent HeLa cells after DOX-HCl treatment. There was no significant difference between the sample of HeLa cells and the sample of HeLa cells with unloaded sperms. After 24 hours, proliferated cells covered the whole substrates of the dishes in both samples. The majority of the cells were in fusiform shape,[51] which suggests that unloaded sperms had no influence on the viability of adherent HeLa cells. In the sample with DOX-HCl-loaded sperms, DOX-HCl was barely found in HeLa cells on the substrate initially because most of the sperms swam freely in the bulk solution. During 72 h of incubation, more and more DOX-HCl was detected around the HeLa cells together with the sperm motility decreasing. After 24 hours, the density of HeLa cells was only half of that of the other two groups. Cells were round in morphology and exhibited blebbing or cytoplasmic extrusions, which indicates cell apoptosis and the termination of cell proliferation.[52] After 72 hours, ruptured membranes and nuclei were observed in the sample. DOX-HCl-coupled nuclei and organelles were stained in red.

Hela spheroids were cultured as three dimensional *in vitro* model of tumor tissue.[53] To avoid the cell apoptosis effect of DOX-HCl, we employed FITC-BSA (Fluorescein isothiocyanate labelled bovine serum albumin) at first as a model drug to observe the drug distribution in spheroids (Figure S4). After 24 hours co-incubation of BSA-loaded sperms with spheroids, sperms were found not only in the solution, but also in the spheroids as shown in the overlaid z-stack images. This proves the tissue penetration capability of sperms. According to a semi-quantitative analysis by ImageJ, the integrated fluorescence intensity which represents the total amount of BSA increased 1.8 times compared to the amount at the beginning. The spreading area of FITC-BSA increased 7.4 times.

This indicates that the increase of the fluorescent area was not only because more sperms penetrated into the spheroid over time, but also because more FITC-BSA was transferred from sperms into the HeLa spheroid.

Cell-killing efficacy was investigated by co-incubation of DOX-HCl-loaded sperms ($8\times10^4$ sperms) with HeLa spheroids. Spheroids without any sperms or drugs, with only unloaded sperms and with only DOX-HCl solution were cultured as control experiments. Figure 4a illustrates the drug transport into a spheroid during 72 h when it was treated with DOX-HCl-loaded sperms. Red fluorescence shows the average intensity of 36 overlaid z-stack images and indicates the presence of DOX-HCl. Gradually, DOX-HCl was found in the center of the spheroid over time. After 72 h, the size of all spheroids decreased owing to drug-induced cell apoptosis. In addition, broken clusters and ruptured cells were observed in the medium (Figure 4a at 72 h). Cell viability analysis was performed by LIVE/DEAD staining method.[54] Fluorescence images of cells of digested spheroids after staining are shown in Figure 4b, in which live cells and dead cells are in green and red color, respectively. After 72 h, the percentage of dead cells after treatment with DOX-HCl-loaded sperms was higher than in the control samples. Quantitative results of cell counting are shown in Figure 4d. In the first 24 h of culture, there was no significant change in all groups, while after 48 h, DOX-HCl-loaded sperms showed a cell-killing *effect* comparable to the treatment with DOX-HCl solution with the same amount of DOX-HCl as the one loaded into the sperm cells (1.5 µg). A lower percentage of live cells was found after the treatment with DOX-HCl-loaded sperms (47%) and DOX-HCl solution (45%) compared to the spheroid control group (68%). The group with DOX-HCl-loaded sperms showed the lowest percentage of live cells (13%) among all groups after 72 h. Unloaded sperms showed a negative effect on HeLa spheroids as well, as the percentage of live cells was only 37%, attributed to the spheroid disintegration induced by the sperm beating and hyaluronidases reaction (from sperm membrane) with the extracellular matrix. No significant decrease of cell viability was found in the other two groups between 48 and 72 h.

During cell culture in the lab, HeLa spheroids sustain a balance between cell proliferation and apoptosis.[55] When the number of live cells increases, the number of dead cells also increases, especially in the necrotic core of the spheroid as the nutrients can hardly reach the interior cells. In our experiment, unloaded sperms showed no influence on HeLa cells in the first 48 h of treatment. The significantly improved cell-killing effect that we observed after 72 h on HeLa spheroids compared to the adherent HeLa cells can be explained by the penetration of motile sperms which results in spheroid disaggregation. This effect of the sperm cells might be attributed to the vigorous sperm tail beating and a sperm-membrane enzyme reaction with extracellular hyaluronic acid, a ubiquitous carbohydrate polymer that is part of the extracellular matrix in tumor tissue. DOX-HCl in solution phase can be rapidly taken in by the spheroid's outer cell layer. Consequently, a pronounced effect of the DOX-HCl solution group in the first 48 h was observed. However, when dissipated in the cell medium, diluted DOX-HCl was apparently not sufficient to induce apoptosis of more cells from 48 h to 72 h (Figure 4d). This manifests an advantage of the sperm-hybrid delivery system in an *in vivo* application scenario: the ability to avoid drug dilution in body fluids. While protecting the drug by the cell membrane, sperm cells transport it through the diffusion barrier of the tumor spheroid with their motility and cell-fusion ability. Overall, by means of functional combination of cargo protection and tissue penetration of sperms and the cell-killing efficacy of the drug, this biocompatible delivery strategy employing DOX-HCl-loaded sperm cells poses an intriguing alternative dosage form that is able to induce a death rate of nearly 90% after 72 h of treatment on HeLa spheroids, as was shown in Figure 4d.

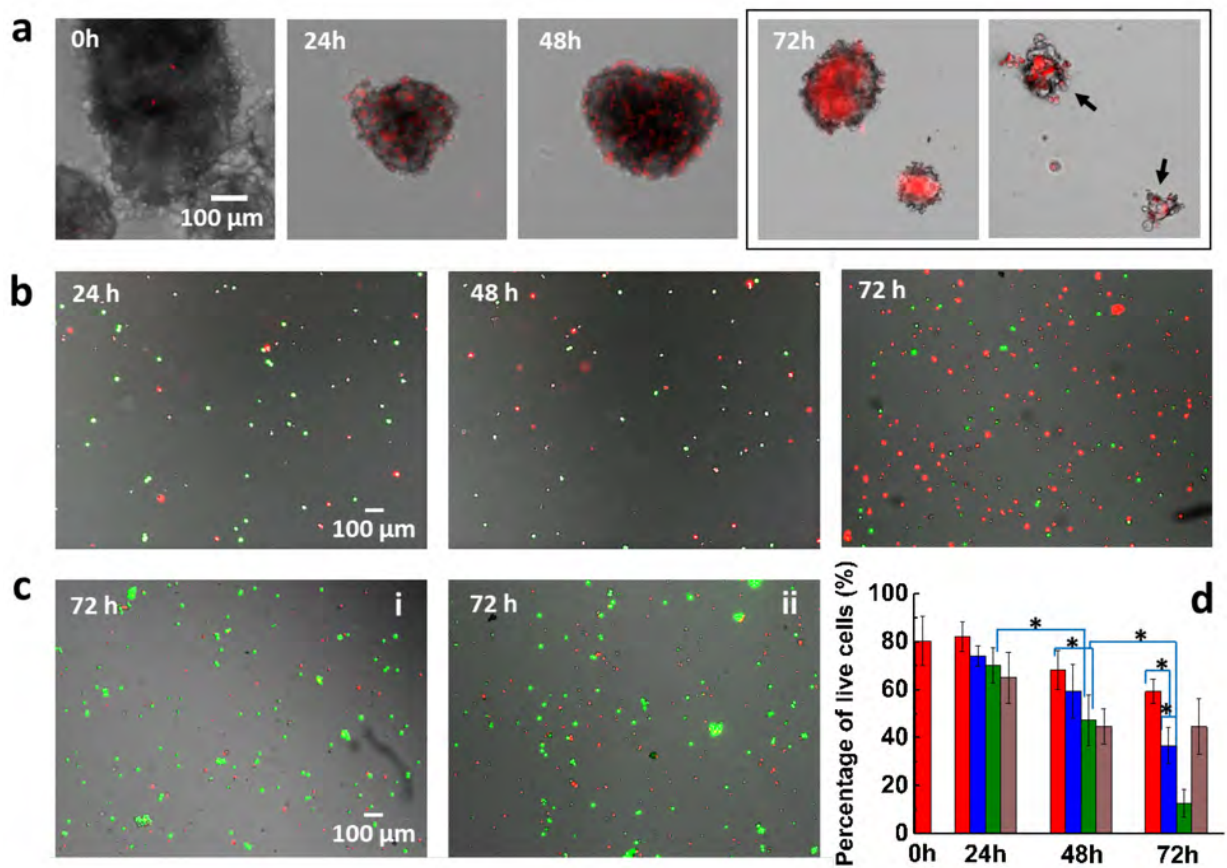

**Figure 4.** Cell-killing effect of DOX-HCl-loaded sperms on HeLa spheroids. (a) Overlaid z-stack images of HeLa spheroids under treatment by DOX-HCl-loaded sperms. Red color shows the fluorescence of DOX-HCl under an excitation light with a wavelength of 470 nm. Black arrows point at ruptured spheroids. (b) LIVE/DEAD staining images of cells of digested spheroids from the group of spheroids with DOX-HCl-loaded sperms treatment. (c) Live/dead staining images of cells of digested spheroids from the group of control spheroids after 72 h: (i) spheroids without sperms and (ii) spheroids with unloaded sperms. (d) Histogram of the portion of live cells relative to the total amount of cells at different time points. Red columns represent control spheroid, blue columns represent unloaded sperm treatment, green columns represent DOX-HCl solution treatment and brown columns the DOX-loaded sperm treatment (Count = $10^4$ for each group, * $p < 0.01$, ANOVA analysis).

A complete *in vitro* targeted drug delivery experiment was performed by means of a microfluidic channel to mimic *in vivo* drug administration by sperm-hybrid micromotors (Figure S5). DOX-HCl-loaded sperms were prepared as mentioned before (detailed information in SI). Considering the auto-fluorescence of DOX-HCl, a fluorescence microscopy video was taken under illumination of a 470 nm laser in order to check the motility of drug-loaded sperms (Video S5). Although immotile sperms tend to agglomerate, both motile and immotile sperms were able to encapsulate the drug. To clearly observe the details of the sperm release process on cells, sperm-hybrid micromotors were guided first to HeLa clusters instead of dense spheroids (Figure 5a, Video S6). Here, the coupled sperm cell swam into the cell cluster after being released, and then the sperm head connected to the cells in the cluster due to membrane adhesion. Videos S7 and S8 display FITC-BSA- and DOX-HCl-loaded sperm cell transport through the constriction channel and sperm release onto a tumor spheroid, respectively. The whole journey path was around 2 cm long and the journey took 8 minutes. The sperm was released into the spheroid when the tetrapod arms hit the outer boundary of the tumor spheroid, and then continued swimming into the spheroid until it was trapped inside. Figure 5b shows the distribution of released DOX-HCl within the spheroid over time. The fluorescence intensity at the sperm position decreased while the fluorescent area within the spheroid increased. This indicates that DOX-HCl was released from the sperm and distributed within the spheroid. However, the integrated fluorescence intensity of the whole sperm-spheroid complex increased by two times over time, probably due to unanticipated DOX-HCl release from other nearby sperms (Figure S6). Although the microfluidic channel was designed to separate the dosing region (where the drug-loaded sperm is coupled to the micromotor) from the disease region (HeLa cell spheroid), it was not possible to completely avoid that active sperms unintentionally reached the spheroid. Nonetheless, a number of cells were found with apoptotic blebs after 24 hours. Figure 5c shows the membrane fusion of a sperm head and a HeLa cell in 24 hours after sperm release, the anterior part of the sperm head was not visible anymore while the posterior part of the sperm head and the flagellum could still be distinguished. Previous research explained this interaction with the somatic fusion ability of acrosome-reacted spermatozoa.[27] Taking advantage of the cell fusion ability of sperms, our sperm-hybrid system shows a practical potential to enhance the drug uptake and availability by transporting it from cell to cell (sperm to HeLa cell) without dilution into the extracellular medium.

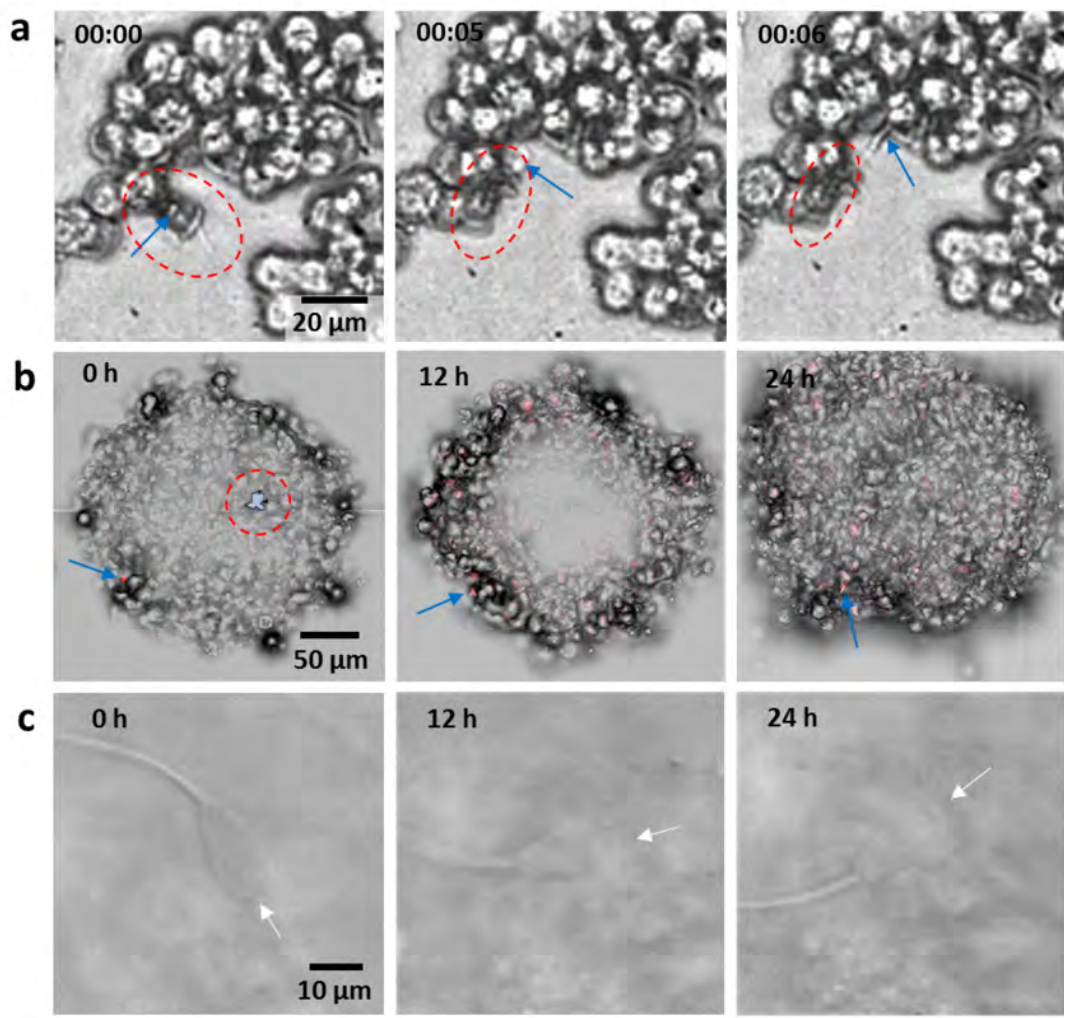

**Figure 5.** (a) Image sequence of the sperm release process when the arms hit HeLa cells. Time lapse in min:s. (b) Overlaid z-stack images of DOX-HCl distribution in a HeLa spheroid. (20 images with a stack separation distance of 5 µm) (c) Overlaid z-stack images of sperm-HeLa cell fusion. (15 images with a stack separation distance of 1 µm) Blue arrows point at the positions of the sperm head. Red dashes circle the colored tetrapods. White arrows point at the sperm membrane.

CONCLUSIONS

We have proposed a novel drug delivery system based on sperm-hybrid micromotors. In such assembly, sperms are utilized as drug carriers for potential cancer treatment in the female reproductive tract, as the sperm membrane can penetrate cancer cells thanks to its capacity to fuse with somatic cells,[56] efficiently transferring the drug to the target cell/model tumor in the process. Moreover, the sperm cells serve as propulsion source while the magnetic microstructure is used for guidance and release of the sperm: When the arms of the microstructure hit Hela cells, they bend and thus open a way to free the sperm. Bovine sperms were used as model cells to load DOX-HCl drug for treatment of cervical cancer. For that, HeLa cell spheroids were cultured as an *in vitro* tumor model. DOX-HCl was locally distributed into the HeLa cell clusters after the sperm cells were released, showing higher tumor cell-killing efficacy within the first 48h, compared to the drug solution with the same dose. Sperms also showed surprisingly high drug encapsulation capacity compared to liposomes or other synthetic carriers. Furthermore, the sperm was capable to swim through complex environments in an efficient manner not only due to their tail beating but also due to their membrane biochemistry. A set of enzymes is expressed by sperms to catalyze the degradation of hyaluronic acid, which contributes to the constitution of the extracellular matrix of oocyte-surrounding cumulus cells.[57] It was widely reported that hyaluronic acid also plays an important role in the proliferation and migration of tumor tissue.[58] Therefore the motility and the hyaluronidases wielded by sperm cells allow them to penetrate deep into a tumor spheroid for effective *in situ* drug administration. Besides, sperm cells can remain functional in the human body for a longer time in comparison to other foreign cells due to the ability to inhibit the immune response by displaying specific glycans on the membrane.[59] This reduces undesired immune-response and thus makes this system compatible to the host body. Such sperm-hybrid micromotors not only have potential application for gynecologic cancer treatment but also for treating other diseases in the female reproductive tract such as endometriosis, pelvic inflammatory diseases, among others. Such devices can be also engineered to carry genes, mRNA, imaging contrast agents, among other substances of interest for diverse biomedical applications. Furthermore, these self-propelled carriers are promising to avoid the dilution in body fluids and undesired accumulation of such cargo, in contrast to other conventional carriers, as the cargo is internalized by the sperm and protected by its membrane. This drug loading process does not seem to interfere with sperm motility due to the sperm cell's incomplete metabolism that avoids intracellular degradation of the drug. Although there are still some challenges to overcome before this system can be applied in *in vivo* environments (e.g. imaging, biodegradation of the synthetic part, multiple sperms carrying and delivery, and improved control of sperm release), sperm-hybrid systems may be envisioned to be applied in *in situ* cancer diagnosis and treatment in the near future.

ASSOCIATED CONTENT

AUTHOR INFORMATION

Corresponding Author

Dr. Mariana Medina Sánchez

m.medina.sanchez@ifw-dresden.de

Present Address

Institute for Integrative Nanosciences, IFW Dresden, Helmholtzstraße 20, 01069 Dresden

Author Contributions

H.X, M.M.-S and O.G.S. conceived the project. H.X and M.M.-S designed the experiments with help from V.M. and F.H. H.X. performed the experiments. H.X., M.M.-S and V.M. analyzed the results. H.X., M.M.-S wrote the manuscript. L.S. helped with the fabrication. All authors commented and edited the manuscript and figures. All authors have given approval to the final version of the manuscript.

Funding Sources

Chinese Scholarship Council (CSC)

Notes

The authors declare no competing financial interest

ACKNOWLEDGMENT

The authors thank Masterrind GmbH for kind donation of cryopreserved bovine semen. We thank Stefan Harazim, Martin Bauer, Ronny Engelhard, Sandra Nestler and Cornelia Krien for clean room support. We thank Dr. Daniil Karnaushenko for the support on elasticity measurement and Shunyao Zhang for his help on the simulation work. Thanks to Maria Guix, Sonja Weiz and Britta Koch for their helpful discussions.REFERENCES

(1) Cho, K.; Wang, X.; Nie, S.; Chen, Z. G.; Shin, D. M. *Clinical cancer research : an official journal of the American Association for Cancer Research* **2008,** 14, (5), 1310-6.
(2) De Jong, W. H.; Borm, P. J. *International journal of nanomedicine* **2008,** 3, (2), 133-49.
(3) Yingchoncharoen, P.; Kalinowski, D. S.; Richardson, D. R. *Pharmacological Reviews* **2016,** 68, (3), 701-787.
(4) Shi, J.; Votruba, A. R.; Farokhzad, O. C.; Langer, R. *Nano Lett* **2010,** 10, (9), 3223-30.
(5) Bertrand, N.; Wu, J.; Xu, X.; Kamaly, N.; Farokhzad, O. C. *Adv Drug Deliv Rev* **2014,** 66, 2-25.
(6) Allen, T. M.; Cullis, P. R. *Advanced Drug Delivery Reviews* **2013,** 65, (1), 36-48.
(7) Tannock, I. F.; Lee, C. M.; Tunggal, J. K.; Cowan, D. S.; Egorin, M. J. *Clinical cancer research : an official journal of the American Association for Cancer Research* **2002,** 8, (3), 878-84.
(8) Allen, T. M.; Hansen, C. B.; de Menezes, D. E. L. *Advanced Drug Delivery Reviews* **1995,** 16, (2–3), 267-284.
(9) Tan, S.; Wu, T.; Zhang, D.; Zhang, Z. *Theranostics* **2015,** 5, (8), 863-81.
(10) Stuckey, D. W.; Shah, K. *Nature Reviews Cancer* **2014,** 14, (10), 683-691.
(11) Xuan, M.; Shao, J.; Dai, L.; Li, J.; He, Q. *ACS applied materials & interfaces* **2016,** 8, (15), 9610-9618.
(12) Hamidi, M.; Zarrin, A.; Foroozesh, M.; Mohammadi-Samani, S. *Journal of controlled release* **2007,** 118, (2), 145-160.
(13) Magdanz, V.; Sanchez, S.; Schmidt, O. G. *Advanced Materials* **2013,** 25, (45), 6581-6588.
(14) Tanaka, Y.; Morishima, K.; Shimizu, T.; Kikuchi, A.; Yamato, M.; Okano, T.; Kitamori, T. *Lab on a Chip* **2006,** 6, (3), 362-368.
(15) Hosseinidoust, Z.; Mostaghaci, B.; Yasa, O.; Park, B.-W.; Singh, A. V.; Sitti, M. *Advanced Drug Delivery Reviews* **2016,** 106, Part A, 27-44.
(16) Zhuang, J.; Sitti, M. *Scientific Reports* **2016,** 6, 32135.
(17) Felfoul, O.; Mohammadi, M.; Taherkhani, S.; De Lanauze, D.; Xu, Y. Z.; Loghin, D.; Essa, S.; Jancik, S.; Houle, D.; Lafleur, M. *Nature Nanotechnology* **2016**.
(18) Akin, D.; Sturgis, J.; Ragheb, K.; Sherman, D.; Burkholder, K.; Robinson, J. P.; Bhunia, A. K.; Mohammed, S.; Bashir, R. *Nat Nano* **2007,** 2, (7), 441-449.
(19) Palffy, R.; Gardlik, R.; Hodosy, J.; Behuliak, M.; Resko, P.; Radvansky, J.; Celec, P. *Gene Ther* **2005,** 13, (2), 101-105.23

(20) Siegel, R. L.; Miller, K. D.; Jemal, A. *CA: A Cancer Journal for Clinicians* **2016,** 66, (1), 7-30.
(21) Cummins, J.; Woodall, P. *Journal of Reproduction and Fertility* **1985,** 75, (1), 153-175.
(22) Johnson, G. D.; Lalancette, C.; Linnemann, A. K.; Leduc, F.; Boissonneault, G.; Krawetz, S. A. *Reproduction (Cambridge, England)* **2011,** 141, (1), 21-36.
(23) Kwon, G. S.; Okano, T. *Advanced Drug Delivery Reviews* **1996,** 21, (2), 107-116.
(24) Srivastava, S. K.; Medina-Sánchez, M.; Koch, B.; Schmidt, O. G. *Advanced Materials* **2016,** 28, (5), 832-837.
(25) Wu, Z.; Lin, X.; Wu, Y.; Si, T.; Sun, J.; He, Q. *ACS Nano* **2014,** 8, (6), 6097-6105.
(26) Bysell, H.; Månsson, R.; Hansson, P.; Malmsten, M. *Advanced Drug Delivery Reviews* **2011,** 63, (13), 1172-1185.
(27) Mattioli, M.; Gloria, A.; Mauro, A.; Gioia, L.; Barboni, B. *Reproduction (Cambridge, England)* **2009,** 138, (4), 679-87.
(28) Medina-Sánchez, M.; Schwarz, L.; Meyer, A. K.; Hebenstreit, F.; Schmidt, O. G. *Nano letters* **2015,** 16, (1), 555-561.
(29) Gravance, C. G.; Vishwanath, R.; Pitt, C.; Garner, D. L.; Casey, P. J. *Journal of Andrology* **1998,** 19, (6), 704-709.
(30) Van Dilla, M. A.; Gledhill, B. L.; Lake, S.; Dean, P. N.; Gray, J. W.; Kachel, V.; Barlogie, B.; Gohde, W. *The journal of histochemistry and cytochemistry : official journal of the Histochemistry Society* **1977,** 25, (7), 763-73.
(31) Ishijima, S. *Reproduction (Cambridge, England)* **2011,** 142, (3), 409-415.
(32) Ishimoto, K.; Gaffney, E. A. *Journal of The Royal Society Interface* **2016,** 13, (124), 20160633.
(33) Armon, L.; Eisenbach, M. *PLOS ONE* **2011,** 6, (12), e28359.
(34) Magdanz, V.; Medina-Sánchez, M.; Chen, Y.; Guix, M.; Schmidt, O. G. *Advanced Functional Materials* **2015,** 25, (18), 2763-2770.
(35) *Annual Review of Fluid Mechanics* **2011,** 43, (1), 501-528.
(36) Jikeli, J. F.; Alvarez, L.; Friedrich, B. M.; Wilson, L. G.; Pascal, R.; Colin, R.; Pichlo, M.; Rennhack, A.; Brenker, C.; Kaupp, U. B. *Nature Communications* **2015,** 6, 7985.
(37) Frimat, J. P.; Bronkhorst, M.; de Wagenaar, B.; Bomer, J. G.; van der Heijden, F.; van den Berg, A.; Segerink, L. I. *Lab on a Chip* **2014,** 14, (15), 2635-2641.
(38) Sun, H.-B.; Kawata, S., Two-photon photopolymerization and 3D lithographic microfabrication. In *NMR• 3D Analysis• Photopolymerization*, Springer: 2004; pp 169-273.
(39) Ha, C. W.; Yang, D.-Y. *Optics express* **2014,** 22, (17), 20789-20797.
(40) Nosrati, R.; Graham, P. J.; Liu, Q.; Sinton, D. *Scientific Reports* **2016,** 6, 26669.
(41) Tacar, O.; Sriamornsak, P.; Dass, C. R. *Journal of Pharmacy and Pharmacology* **2013,** 65, (2), 157-170.
(42) Ishida, T.; Atobe, K.; Wang, X.; Kiwada, H. *Journal of controlled release* **2006,** 115, (3), 251-258.
(43) Makhluf, S. B.-D.; Abu-Mukh, R.; Rubinstein, S.; Breitbart, H.; Gedanken, A. *Small* **2008,** 4, (9), 1453-1458.
(44) Rengan, A. K.; Agarwal, A.; van der Linde, M.; du Plessis, S. S. *Reproductive Biology and Endocrinology* **2012,** 10, (1), 92.
(45) Kiyomiya, K.-i.; Matsuo, S.; Kurebe, M. *The Mode of Nuclear Translocation of Adriamycin-Proteasome Complex* **2001,** 61, (6), 2467-2471.

(46) Batrakova, E. V.; Li, S.; Brynskikh, A. M.; Sharma, A. K.; Li, Y.; Boska, M.; Gong, N.; Mosley, R. L.; Alakhov, V. Y.; Gendelman, H. E. *Journal of Controlled Release* **2010,** 143, (3), 290-301.
(47) Gur, Y.; Breitbart, H. *Mol Cell Endocrinol* **2008,** 282, (1-2), 45-55.
(48) Qu, Q.; Ma, X.; Zhao, Y. *Nanoscale* **2015,** 7, (40), 16677-86.
(49) Munerati, M.; Cortesi, R.; Ferrari, D.; Di Virgilio, F.; Nastruzzi, C. *Biochimica et Biophysica Acta (BBA)-Molecular Cell Research* **1994,** 1224, (2), 269-276.
(50) Gill, D. R.; Hyde, S. C.; Higgins, C. F.; Valverde, M. A.; Mintenig, G. M.; Sepúlveda, F. V. *Cell* **1992,** 71, (1), 23-32.
(51) Liao, F.; Hu, Y.; Wu, L.; Tan, H.; Luo, B.; He, Y.; Qiao, Y.; Mo, Q.; Wang, Y.; Zuo, Z. *Oncology reports* **2015,** 33, (4), 1823-1827.
(52) Coleman, M. L.; Sahai, E. A.; Yeo, M.; Bosch, M.; Dewar, A.; Olson, M. F. *Nat Cell Biol* **2001,** 3, (4), 339-345.
(53) Yuan, F. *Nat Biotech* **1997,** 15, (8), 722-723.
(54) Ma, H. L.; Jiang, Q.; Han, S.; Wu, Y.; Cui Tomshine, J.; Wang, D.; Gan, Y.; Zou, G.; Liang, X. J. *Molecular imaging* **2012,** 11, (6), 487-98.
(55) Grimes, D. R.; Kelly, C.; Bloch, K.; Partridge, M. *Journal of The Royal Society Interface* **2014,** 11, (92), 20131124.
(56) Bendich, A.; Borenfreund, E.; Sternberg, S. S. *Science (New York, N.Y.)* **1974,** 183, (4127), 857-9.
(57) Gwatkin, R.; Andersen, O. *Journal of reproduction and fertility* **1973,** 35, (3), 565-567.
(58) Liu, N.; Gao, F.; Han, Z.; Xu, X.; Underhill, C. B.; Zhang, L. *Cancer research* **2001,** 61, (13), 5207-14.
(59) Pang, P. C.; Tissot, B.; Drobnis, E. Z.; Sutovsky, P.; Morris, H. R.; Clark, G. F.; Dell, A. *The Journal of biological chemistry* **2007,** 282, (50), 36593-602.

# Supporting Information

# Sperm-hybrid micromotor for targeted drug delivery

*Haifeng Xu[1], Mariana Medina-Sánchez[1]\*, Veronika Magdanz[1], Lukas Schwarz[1], Franziska Hebenstreit[1] and Oliver G. Schmidt[1,2]*

[1]Institute for Integrative Nanosciences, IFW Dresden, Helmholtzstraße 20, 01069 Dresden, Germany

[2]Material Systems for Nanoelectronics, Chemnitz University of Technology, Reichenhainer Straße 70, 09107 Chemnitz, Germany

# EXPERIMENTAL PART

*Microfluidic platform fabrication*

Microfluidic channels were fabricated by soft lithography. Briefly, a silicon wafer was spin coated with the negative photoresist SU-8 (SÜSS Microtec) and patterned by maskless lithography (µPG 501 Maskless Aligner, Heidelberg Instruments). After that, a mixture of PDMS and curing agent (Dow Corning Corp.) was poured onto the obtained mold and cured at 75 °C for 2h. The channels were designed to be 200 µm deep and 3 cm long. Two reservoirs were punched out at the inlet and the outlet after peeling off the PDMS channel from the mold. Finally, to complete the channel fabrication, the PDMS channel and a cleaned glass substrate were physically bonded after being exposed to oxygen plasma for about 0.5 min. To condition the inner channel surface, channels were filled with Pluronic® F-127 solution (10 µg/mL in DI water) (Sigma-Aldrich, Germany) and incubated for 1 h at 37 °C. Pluronic® F-127 has been previously used to repel sperm cells, making it suitable for minimizing unspecific adhesion between sperms, the channel, and micromotor surfaces.[1] Before performing an experiment, the resulting channels were always rinsed with sperm medium (SP-TALP, see Table S1) 3 times.

*Tetrapod fabrication*

Arrays of polymeric tetrapods were fabricated by 3D laser lithography (Photonic Professional GT, Nanoscribe GmbH). Briefly, dip-in laser lithography photoresist (IP-Dip, Nanoscribe GmbH) was dropped onto cleaned quartz glass substrates (25x25 mm²) as the basis polymer material for laser writing. The negative-tone photoresist was then polymerized specifically at preprogrammed exposure positions by two-photon absorption (laser wavelength 780 nm). The design of the tetrapod arrays was programmed with DeScribe software (Nanoscribe GmbH).

The samples were then dried in a critical point dryer (Autosamdri®-931, Tousimis Research Corporation) after 20 min of development in mr-Dev 600 (Micro Resist Technology GmbH) and 3 min of washing in isopropanol. The dried samples were coated with 10 nm Fe and 2 nm Ti of high purity (99.995 %) by E-beam metal evaporation (Edwards auto 500 e-beam, Moorfield Nanotechnology Limited). To create a magnetic easy axis, the sample holder was tilted at an angle of 75° during the deposition process. The metal-coated samples were also immersed in Pluronic® F-127 solution for 1 h at 37°C and rinsed with water and SP-TALP for subsequent experiments to avoid unspecific adhesion of target cells.

Tetrapod printing quality was evaluated by scanning electron microscopy (Zeiss NVision 40, Carl Zeiss Microscopy GmbH). The samples were fixed on a metal stub and coated with 10 nm of platinum. Imaging was performed at a working distance of 5 mm in the secondary electron imaging mode at a working voltage of 2 kV.

The finite element analysis of the arm deformation was performed with Autodesk Inventor software (applied Young's modulus and poisson ratio are 0.15 GPa and 0.45, respectively)[2]. The relationship between the deflection and the distributed load was calculated according to the Euler–Bernoulli equation[3] as:

$$q = \frac{d^2}{dx^2}\left(EI\frac{d^2\omega}{dx^2}\right),$$

where ω and q represent the deflection and distributed load, respectively, and $x$ describes the orientation direction of an arm and $E$ is the Young's modulus of the material. $I$ is the second moment of area of an arm's cross-section. For an arm with a loading along the $z$ axis,

$$I = \iint z^2 dy\, dz.$$

*Preparation of drug-loaded sperm cells*

Bovine sperm cells were recovered by thawing cryopreserved sperm straws rapidly in a water bath at 38 °C for 2 min, and washed with BoviPure 100 / BoviDilute (40%/ 80%).[4] After 5 min centrifugation at 300 g in soft mode, the sperms were resuspended in 1 mL SP-TALP for subsequent use. Sperm concentration was calculated by using a cell counting chamber. 1 mg/mL of FITC-BSA (Fluorescein isothiocyanate conjugated bovine serum albumin) (Sigma-Aldrich, Germany) and DOX-HCl (doxorubicin hydrochloride) (Sigma-Aldrich, Germany) solution were prepared in SP-TALP, respectively, and stored under dark conditions at 4 °C. To prepare drug-loaded sperm cells, a mixture of sperm solution and FITC-BSA or DOX-HCl solution at specific concentrations were co-incubated in a humidified atmosphere of 5% $CO_2$ in air at 37 °C for 1 h. After washing the sample 2 times with SP-TALP by centrifugation at 300 g for 5 min, the pellet of drug-loaded sperms was redispersed in SP-TALP and stored in the incubator under dark conditions for subsequent use. It is important to note that the samples were to be used within 6 h to guarantee sperm motility.

*HeLa cell spheroid preparation*

HeLa cells were cultured in a 25 mL adherent flask for 2 weeks after recovering (for details of the medium see Table S2). To prepare tumor spheroids, the cells were first incubated in 2 mL Trypsin/EDTA for 10 min to be detached from the substrate. After dilution in 8 mL medium, the cell suspension was then centrifuged at 1000 rpm for 3.5 min to remove the medium. Then the cell pellet was resuspended in 1 mL medium. After calculation, $3\times10^5$ cells were suspended into 3 mL of medium and seeded in a spheroid-culture dish (Cellstar® Cell-Repellent Surface, Greiner bio-one).[5] After 3 days maturation, the spheroid solution was transferred into a 15 mL falcon and incubated for 5 min to separate the sediment cells.

Finally, the bottom pellet of the spheroids was resuspended in 3 mL of new medium in the spheroid-culture dish for subsequent experiments.

*Evaluation of sperm transport procedure*

PDMS-based microfluidic chips were employed to evaluate the sperm-tetrapod coupling procedure, the magnetic guidance of the hybrid micromotors, and the sperm release by the mechanical trigger function. The chip featured a 500 μm wide channel where the sperm-tetrapod coupling and guidance occurred, and obstacles in 3 different shapes where the sperm release occurred (see Video S4). Tetrapods, PDMS channels and sperms were prepared and treated as mentioned before. After that, an array of tetrapods (1296 devices in total) was detached from the substrate by mechanical scratching and dispersed into 50 μL of SP-TALP. For the evaluation of the sperm-tetrapod hybrid micromotor performance, 50 μL of a mixture of equal parts sperm solution ($3\times10^5$ per mL) and tetrapod suspension (ca. 750 in 25 μL SP-TALP) was introduced into the channel. Real time videos of the samples were recorded under the microscope (ZEISS Axio Scope.A1, Carl Zeiss Microscopy GmbH) with a high speed camera (Phantom Miro eX4, Vision Research Inc.) with 30 frames per second. Working temperature was maintained at 38 °C to assert the motility of the respective sperm sample. After coupling, the sperm-hybrid micromotor was guided by simply rotating a permanent magnet and the resulting directional changes were recorded to evaluate the guidance performance. The operating distance between the magnet and the sample was about 10 cm, resulting in an effective magnetic field of roughly 5 mT. Sperm release was evaluated by guiding the coupled motor onto different obstacles.

*Evaluation of the drug loading procedure*

The success of the drug loading procedure was evaluated by the determination of the encapsulation efficiency. DOX-HCl-loaded sperm cells were prepared as mentioned before with a concentration of 3×10$^6$ sperms per mL. The samples of DOX-HCl solution were designed as a series of concentrations with 10, 20, 50, 100 and 200 µg of drug per mL. FITC-BSA-loaded sperms were prepared with FITC-BSA solution with a concentration of 100 µg/mL. After incubation of sperms in these solutions, the respective supernatant was collected after centrifugation and filtered through a 2 µm pore size membrane. The sperms were resuspended after purification as mentioned before. Fluorescence images were taken at an excitation wavelength of 470 nm (DOX-HCl: Ex 470 nm, Em 580 nm;[6] FITC-BSA: Ex 470 nm. Em 509 nm[7]) (Cell Observer, Carl Zeiss Microscopy GmbH). The concentration of DOX-HCl was determined with a fluorescence spectrometer in FL-RL mode (SpectraMax® M2, Molecular Devices, LLC.) and the total weight was calculated. SP-TALP solution was used as blank control for all measurements. The concentration of the residual DOX-HCl was determined by measuring the supernatant of each sample after centrifugation. The encapsulation efficiency was calculated as the ratio of encapsulated drug to the total amount of the used drug:

$$DOX\ loading\ ratio = \frac{total\ weight\ of\ the\ used\ DOX - weight\ of\ residual\ DOX}{total\ weight\ of\ the\ used\ drug}$$

$$DOX\ amount\ per\ sperm = \frac{total\ loading\ amount}{number\ of\ used\ sperms}$$

*Evaluation of drug-loaded sperm delivery toward tumor spheroids*

Figure S6 shows the microfluidic chip that was designed for drug delivery experiments. The chip was designed to have a wide dosing region (where sperms are coupled to the tetrapods), a treatment region (where the tumor spheroid is located) and a constricted neck region in the middle to prevent the entry of non-coupled sperms into the treatment region. Drug-loaded sperms were prepared as mentioned earlier and observed under a fluorescence microscope using an excitation light with a wavelength of 470 nm (see Video S5, exposure time: 800 ms).

After chip treatment with Pluronic® F-127 and SP-TALP, HeLa cell medium was first filled into the chip as the experimental environment. Then the HeLa spheroid was introduced into the chip from the left end and then 5 µL of tetrapod suspension (around 300) and 5 µL of drug-loaded sperms solution ($3\times10^5$ /mL) were introduced from the right end. Real time videos of the transport process were recorded under the aforementioned microscope. With the guidance by an external magnet at a distance of ca. 10 cm, sperm-hybrid micromotors were guided to the target spheroid and the coupled drug-loaded sperms were released into the spheroid. Afterwards, the whole system was incubated in appropriate cell culture conditions (5% $CO_2$ in air at 37 °C). At fixed time points, z-stack images of the sperm embedded spheroids were captured under the aforementioned fluorescence microscope. The exposure time of the fluorescent channel for all images was fixed at 800 ms. Subsequent semiquantitative analysis of the multi-channel images was executed with ImageJ.

**Table S1.** Recipes for Sperm medium and HeLa cell medium

*Sperm-medium (SP-TALP)*

| SP-TL (IVL03)   | Caisson lab   | 9.5 mL |
|-----------------|---------------|--------|
| Albumin (A7030) | Sigma Aldrich | 60 mg  |
| Sodium pyruvate | Gibco         | 0.5 mL |
| Gentamycin      | Cassion lab   | 20 µl  |

*HeLa medium*

| DMEM - high glucose     | Sigma Aldrich | 50 mL  |
|-------------------------|---------------|--------|
| FBS 10%                 | Sigma Aldrich | 5 mL   |
| L-Glutamin 1%           | Gibco         | 0.5 mL |
| Penicillin/Streptomycin | Sigma Aldrich | 0.5 mL |

**SUPPORTING FIGURES**

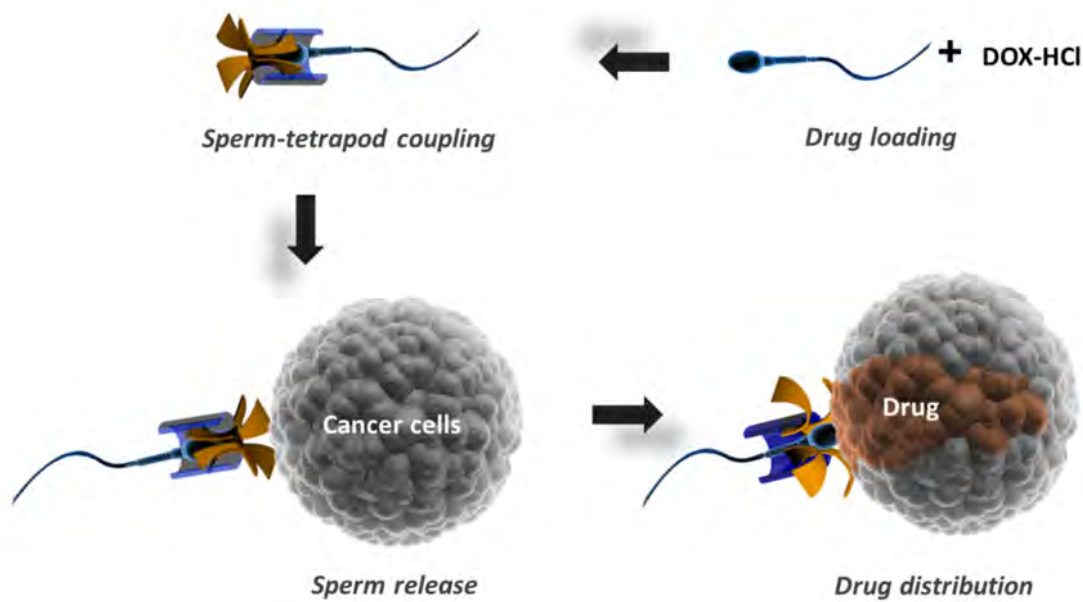

**Figure S1.** Schematic depicting tumor targeted drug delivery by a sperm-hybrid micromotor under magnetic guidance with mechanical sperm release trigger.

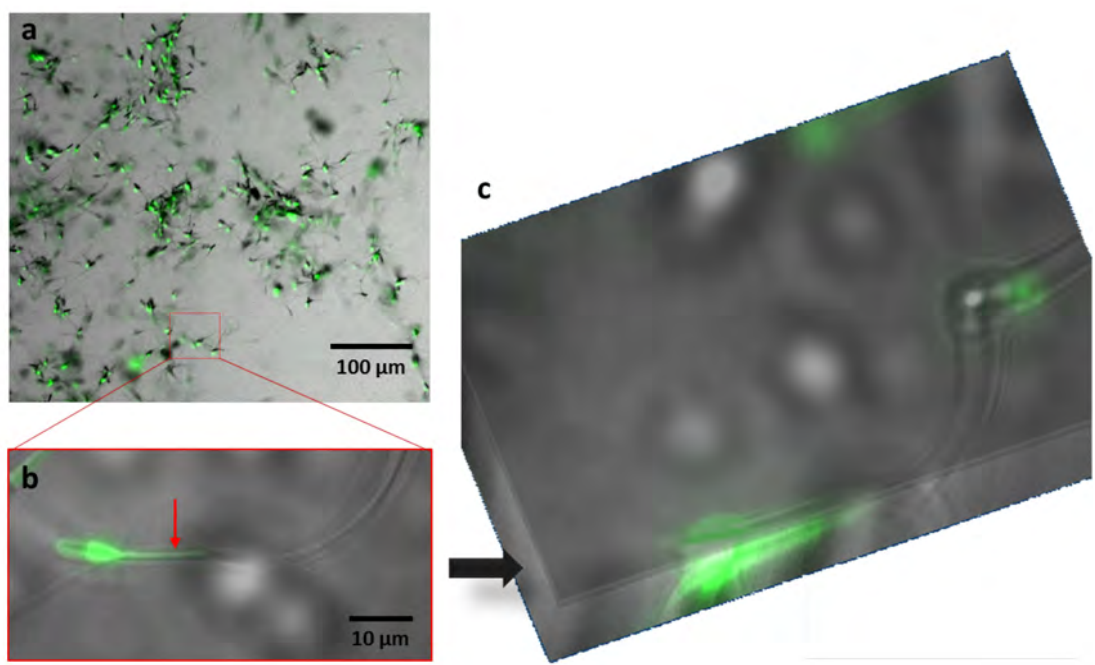

**Figure S2.** Fluorescence and bright field overlaid images of FITC-BSA-loaded sperms in (a) 10×; (b) 40× magnification; (c) 3D reconstruction of 18 z-stack images with stack separation of 0.3 μm showing the cross-section of a BSA-loaded sperm cell.

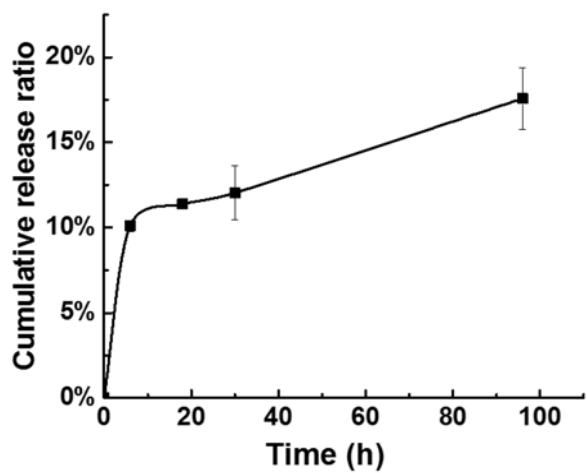

**Figure S3.** Plot of cumulative release ratio of DOX-HCl from loaded sperm over release time. Sperm concentration: $3\times10^6$ /mL. (n = 4)

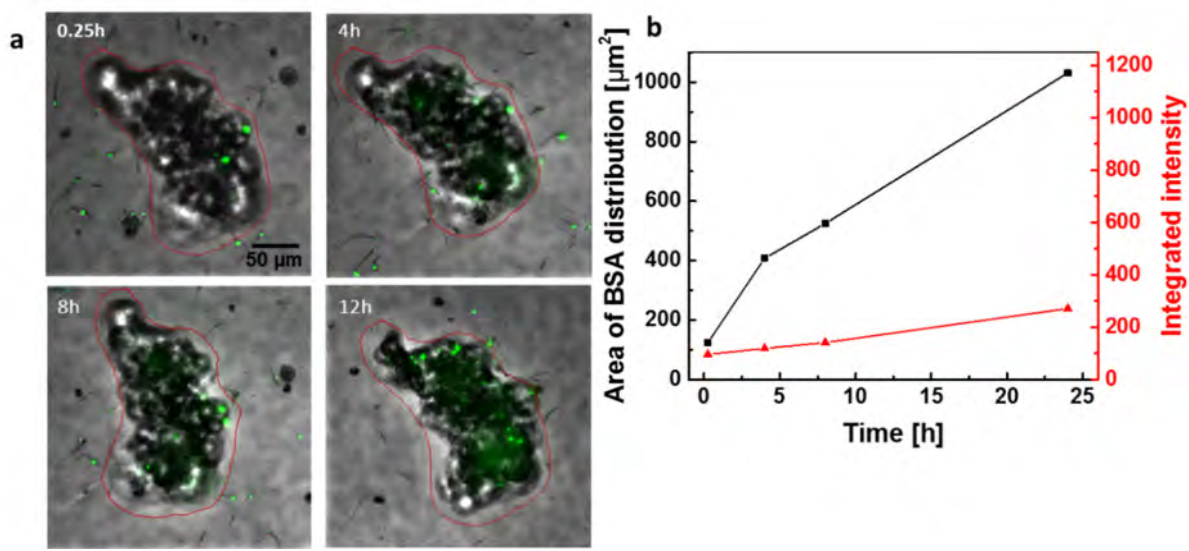

**Figure S4.** FITC-BSA distribution in a HeLa spheroid after the co-incubation of FITC-BSA loaded sperms with a HeLa spheroid. (a) Overlaid z-stack images. Red lines circle the spheroid. FITC-BSA is fluorescing in green under an excitation light of 470 nm. (b) Semiquantitative analysis of the fluorescence intensity of the spheroid shown in (a). Area of BSA distribution describes the spreading area of FITC-BSA fluorescence signals on the spheroid, i.e. drug distribution. Integrated intensity describes the sum of fluorescence intensity in the red-circled spheroid in (a), which corresponds to the total amount of FITC-BSA present.

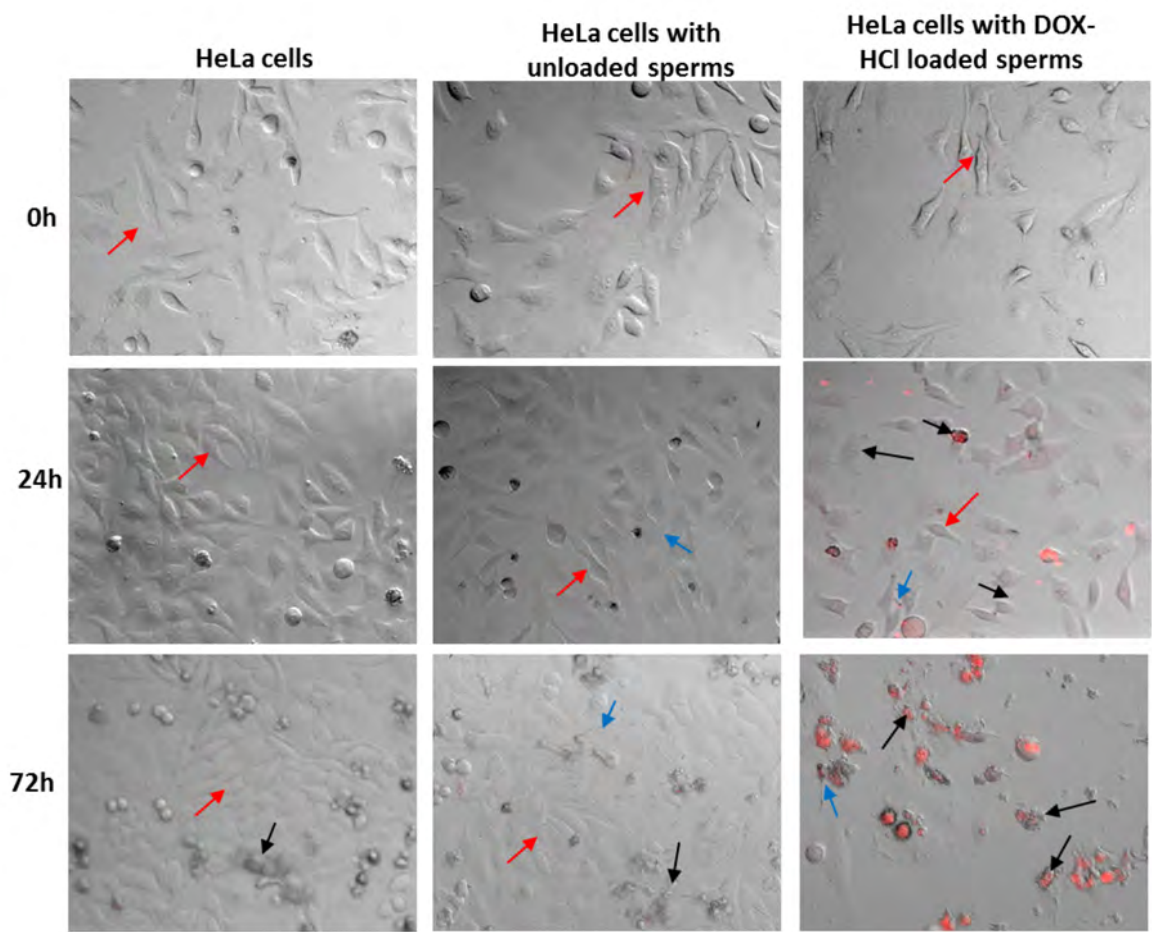

**Figure S5.** Cancer cell apoptosis induced by DOX-HCl-loaded sperms. Red fluorescence indicates DOX-HCl. Red arrows point at proliferated HeLa cells (fusiform). Black arrows point at HeLa cells in apoptosis. Blue arrows point at sperm cells.

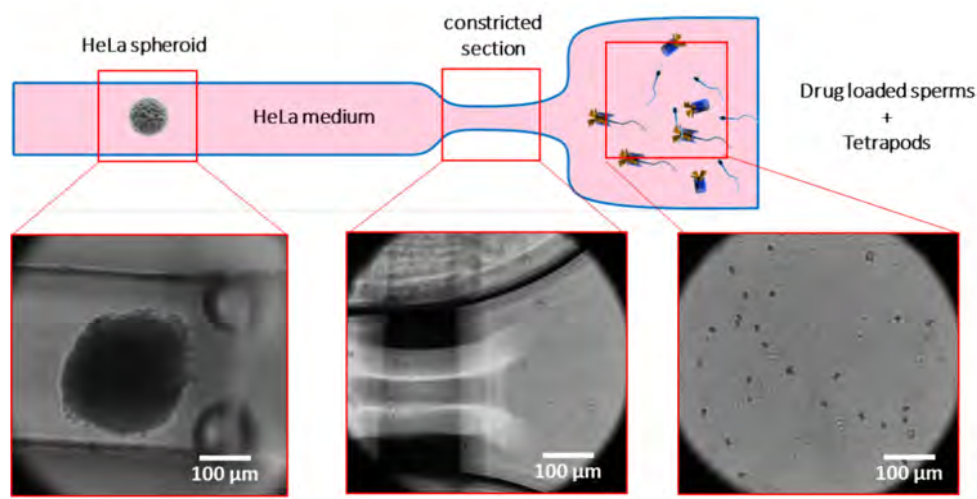

**Figure S6.** Schematic of the microfluidic chip for drug-loaded sperm transport.

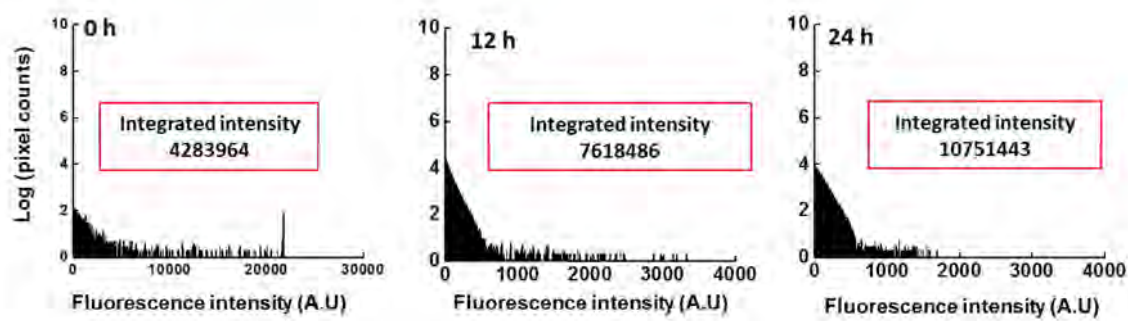

**Figure S7.** Semiquantitative analysis of the DOX-HCl distribution in the HeLa spheroid after the release of the DOX-HCl loaded sperm. Integrated intensity indicates the total amount of DOX-HCl in the sperm and the spheroid.

**REFERENCES**

(1) Frimat, J. P.; Bronkhorst, M.; de Wagenaar, B.; Bomer, J. G.; van der Heijden, F.; van den Berg, A.; Segerink, L. I. *Lab on a Chip* **2014,** 14, (15), 2635-2641.

(2) Bückmann, T.; Kadic, M.; Schittny, R.; Wegener, M. *Physica Status Solidi (b)* **2015,** 252, (7), 1671-1674.

(3) Park, S.; Gao, X. *Journal of Micromechanics and Microengineering* **2006,** 16, (11), 2355.

(4) Samardzija, M.; Karadjole, M.; Getz, I.; Makek, Z.; Cergolj, M.; Dobranic, T. *Reproductive Biology and Endocrinology* **2006,** 4, 58.

(5) Froehlich, K.; Haeger, J.-D.; Heger, J.; Pastuschek, J.; Photini, S. M.; Yan, Y.; Lupp, A.; Pfarrer, C.; Mrowka, R.; Schleußner, E. *Journal of Mammary Gland Biology and Neoplasia* **2016,** 21, (3-4), 89-98.

(6) Kumar, R.; Kulkarni, A.; Nagesha, D. K.; Sridhar, S. *Theranostics* **2012,** 2, (7), 714-722.

(7) Chudakov, D. M.; Verkhusha, V. V.; Staroverov, D. B.; Souslova, E. A.; Lukyanov, S.; Lukyanov, K. A. *Nature Biotechnology* **2004,** 22, (11), 1435-1439.